# A Realisation of Channel Emulation in a Reverberation Chamber method for Over-the-Air Compliance Testing in Support of 3GPP Standardisation

Yunsong Gui, and Tian Hong Loh

[1] Electromagnetic & Electrochemical Technologies, National Physical Laboratory, Teddington, United Kingdom,
yunsong.gui@npl.co.uk, tian.loh@npl.co.uk*

*Abstract*—The inherent long decay power delay profile (PDP) in the reverberation chamber (RC) is a major challenge for accurate channel emulation of 3GPP channel model, which is widely used in performance test of the physical layer. To tackle this challenge, we propose in this paper a novel two-step "closed-loop" approach consisting of (i) a channel measuring step and (ii) a channel model synthesis step. The channel measurement step is used to capture the wireless channel of the RC. In the channel model synthesis step, an additional IQ signal convolution process is introduced prior the IQ signal passes through the channel emulator (CE). This process filters the IQ signal by an equalizer filter derived from the measured channel impulse response (CIR) of the RC obtained in channel measurement step. From the measurement results, the proposed approach is proven that able to effectively emulate typical 3GPP 5G channel model.

## I. INTRODUCTION

As the fifth-generation (5G) and beyond communication technology is being widely commercialized, and the utilized radio frequency (RF) band is being extended to the millimeter-waves (mm-waves), the development of multiple-input-multiple-output (MIMO) over-the-air (OTA) measurement metrology is increasingly becoming important since traditional cable-based measurement methods become obsolete to the most mm-wave RF front ends due to technology development trend over their antenna and RF chain being fully integrated. In recent years, two promising OTA methods have been proposed as candidate protocol-compliance measurement technologies for RF performance testing, namely, the multi-probe anechoic chamber (MPAC) and the radiated two-stage (RTS) method. The former method mimics a dynamic real-world wireless propagation environment based on multiple probes located in an anechoic chamber, aided with MIMO channel emulator (CE) that applicable to emulate most of the test propagation channel scenario. However, the high cost and complex implementation of the test system limits its uptake. The latter method extends the conventional cable-based approach to OTA testing by introducing an OTA based two-stage measurement in the anechoic chamber, with the limitation of being highly dependent on prior knowledge of the antenna radiation pattern of the device under test (DUT) [1-4].

Reverberation chambers (RCs) and the associated metrological methods are powerful tools to perform OTA based measurements for antenna [5] and wireless link performance evaluations. Because of their inherent multipath rich characteristics, RCs are very effective in mimicking indoor wireless propagation environments. Compared with the two types of methods mentioned above, RCs offer advantages over high measurement effectiveness and low implementation cost. However, one of the main drawbacks is their inherent long decay power delay profile (PDP) characteristic, which limits the capability to simulate discrete delay distributed multipath statistic models, such as 5G channel model defined for the 3GPP 5G new radio (NR) protocol [6], even though the RC's PDP can be modified to some extent through modifying the RC quality factor by loading it with absorbing materials [4].

In recent years, researchers proposed various approaches to overcome this difficulty [4], [7-8]. These approaches basically have the generic testbed setup of OTA measurement in RC which includes a CE to simulate the protocol-compliance multipaths channel fading, and a RC to build a RF propagation environment around the DUT. An antenna is used within the RC and is connected to the output of the CE to transmit the RF signal from the CE to the DUT. Therefore, the overall system response of the RC+CE is equal to the convolution of CE's discrete multi-tap response (e.g., 3GPP 5G channel model) with RC's exponential decay response, which drives the needs to research for methods to eliminate the undesired RC's response from the overall system response, and finally achieve the 3GPP 5G channel model.

In [7], instead of assuming the impulse shaping in each tap, a software based modified spatial channel model extended (SCME) model [9] with certain length of delay spread per tap was introduced to the CE. The SCME model was widely used in the fourth generation (4G) wireless network era and has s similar but smaller number of multipath discrete multitap responses to the 3GPP 5G channel model. The similarity between the synthesized channel response measured by a channel sounder and the SCME channel model was evaluated in the paper. While optimized by changing the length of delay spread, the unignored residual PDP of RC in the synthesized channel response limited the performance. In [4], a modified channel simulator scheme based on introducing an extra artificial multipath pattern according to the standard channel model was proposed. Compared with the real taps of standard channel models, the artificial multipaths pose nearly equivalent delay, envelope and 180-degree phase shift. Exploiting the high correlation between every pair of real path and artificial path, the PDP of the RC with continuous exponential decay distribution could be cancelled effectively. However, the effectiveness for the implementation of the CE can be constrained by the hardware limitation, such as the channel sampling rate. In contrast with the open-loop solution which doesn't generate the synthesized channel model based on the measured channel response of the RC, in [8] a closed-loop solution is proposed which consists of two steps namely (i) a channel measuring step and (ii) a channel model synthesis step. However, no practical implementation or measurements were performed in the paper.

In this paper, we present a complete realization of the two-step closed-loop method and evaluate its effectiveness based on experimental measurements in the RC. The paper is organized as follow: Section 2 presents some results on the limitations of the artificial multipath method and a description of the proposed two-step approach. Section 3 provides details about the measurement setup and implementation, and presents measurement results and analysis. Section 4 provides the conclusions.

## II. TWO-STEP CLOSED LOOP METHOD

### A. Assessment of the artificial path method

According to [4], the artificial multipaths are introduced by convolving theoretical SCME taps with $\delta(t) - e^{-\partial t/\tau_{rms}}\delta(t - \partial t)$. The parameter $\partial t$ is the delay interval between the introduced artificial paths and the true SCME taps. The parameter $\tau_{rms}$ is the RMS delay spread of the RC channel response.

The effectiveness of the cancellation can be impressively high under the assumption of the introduced artificial path with a very short delay value to the true path, such as 0.1 ns. However, because of the limitation of the CE implementation, minimum available delay interval between the true path and artificial path is usually limited by the maximum available channel sampling rate and RF bandwidth of the CE.

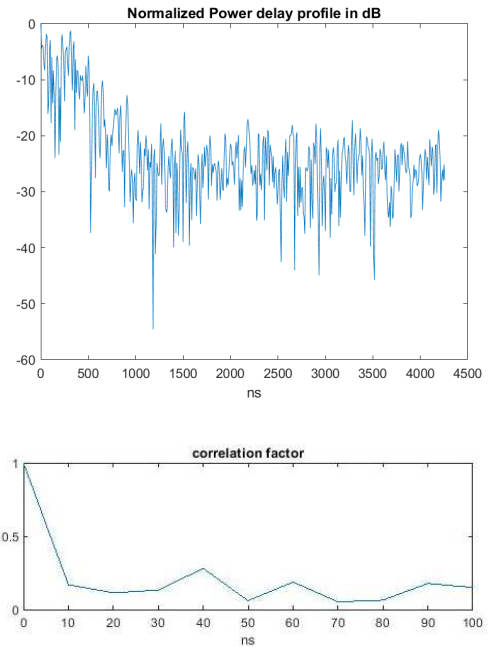

Fig. 1. (1a, Top) Nomalized Power delay profile of the RC after adding artificial path; (1b, Bottom) The self-correlation function of the RC's CIR vs delay interval.

From Fig. 1a, we can see that, if the channel sampling rate of the channel emulator is 100 MS/s, which results in a minimum delay interval between the real paths and the artificial paths of 10 ns, the effectiveness of the cancellation of RC's response is low. It leads to the significant RC's channel response remains in the synthesized channel model. This can also be seen from the self-correlation factor of the RC's channel impulse response in Fig. 1b. The self-correlation factor significantly reduces as the delay interval between the artificial path and the real path increases.

### B. Two-step closed-loop method description

The workflow of two-step closed-loop is shown in the Fig. 2. During the channel measurement step, an equalizer filter can be estimated based on the channel measurement results when the RC channel response changes due to the movement of the stirrer.

The actual RF measurement is executed in the channel model synthesis step, whereby the protocol compliant RF signal goes through a convolution module which convolving the IQ signal by an equalizer filter achieved in the channel measuring step before it enters CE module.

The CE introduces the multipath fading and doppler spread according to the channel model defined in 3GPP specification. The output signal of the CE then goes to the RC, and finally to the receiver where performance statistics are performed.

The target channel model (SCME model or 3GPP 5G channel) is synthesized as the following equations shown.

$$h^X(t) \otimes h^{CE}(t) \otimes h^{RC}(t) = h^{SCME}(t) \qquad (1)$$

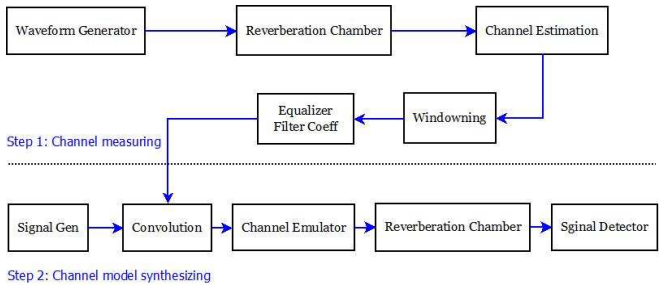

Fig. 2. the workflow of the two-step closed loop method

The channel response of the synthesized channel model $h^{SCME}(t)$, equals to the convolution of the channel response of the RC $h^{RC}(t)$ with the channel responses of a CE $h^{CE}(t)$, and a derived equalized filter $h^X(t)$.

As $h^{SCME}(t)$ can be simulated by setting appropriate parameters of taps according to standard channel model in the CE, i.e. $h^{CE}(t) = h^{SCME}(t)$. This means,

$$h^X(t) \otimes h^{RC}(t) = 1 \qquad (2)$$

So, the $h^X(t)$ is the equalizer filter of the RC's channel impulse response (CIR). In the channel measuring step, the time-domain channel sounder captures snapshot of the RC's wireless propagation channel and estimates the accurate time domain CIR. By utilizing a windowing technique on the estimated CIR, whose window length is based on the maximum delay of multipath of the RC's PDP, the impact of the noise on the CIR estimation can be significantly reduced.

$$h'_{RC} = windowing(h_{RC,CIR}) \qquad (3)$$

According to (2), based on the estimated time domain response of the RC, an equalizer filter can be derived by calculating equalizer filter coefficients with the help of domain transformation as shown in (4). "∗" denotes the complex conjugate.

$$h^X(t) = IFFT\{FFT(h'_{RC}) \times FFT(h'_{RC})^* / |FFT(h'_{RC})|^2\} \qquad (4)$$

## III. MEASUREMENT SETUP AND RESULTS

In this paper, all the measurements were performed within the RC of the UK National Physical Laboratory (NPL).

### A. Channel sounding measurement

To characterize the propagation channel between transmitter and receiver in the RC, a high accuracy channel sounder and the related estimation algorithm were developed. Fig. 3 shows the block diagram for realization of the channel sounder system. As depicted in **Error! Reference source not found.**, the channel sounder system was designed based on a software-defined-radio (SDR) platform. In this work, a pair of National Instrument (NI) vector signal transceivers (VSTs) PXIe-5840 are used as the RF front ends at both transmitting (Tx) and receiving (Rx) ends respectively, with up to 1 GHz RF bandwidth to provide high accurate delay estimation. Meanwhile, two field programmable gate array (FPGA) based baseband processing units have been used as signal generator at Tx end and data streaming at Rx end.

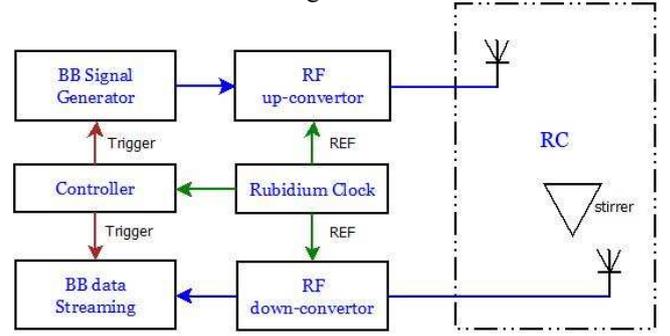

Fig. 3. Block diagram for the channel sounder system and setup in RC

To enable synchronization of the measurement platform, a rubidium clock unit was employed as the reference signal source which provide both Tx and Rx highly stable 10 MHz reference signal with less than 1 ps of timing error accuracy. The reference sharing scheme between the Tx and Rx combined with a hardware trigger in the controller provides the higher synchronization and very low jitter between transmitter and receiver. At the baseband of the Tx end, a Pseudo-Noise (PN) sequence of length 4095 is firstly up-sampled to a two-fold sampling rate and passes through the root raised cosine (RRC) pulse shaping filter to achieve multipath delay resolution to less than 20 ns for the channel estimation of the RC channel. At the Rx end, a post-processing module is developed to estimate the CIR of the RC from the received streamed IQ data sequence. The post-processing module includes an 8190-points length slide correlation module, a windowing module, and a calculating module of the equalizer filter.

The time-domain slide correlation was conducted to achieve the CIR of the RC and a square window was used to significantly reduce the impacts from the noise sample within the CIR. Finally, an equalizer filter was estimated by performing calculations with the help of domain transformation and was output to the channel model synthesis step. In addition, a cable-based calibration procedure was performed at the beginning of each measurement, allowing the effects of the Tx and Rx system response to be measured and compensated based on the calibration results in the postprocessing stage. A pair of directional linearly polarized antennas were used at both Tx and Rx ends. TABLE I. lists the key system settings and parameters.

### B. Setup of reverberation chamber

Fig. 4 shows a photograph of the measurement setup inside the RC. The dimension of the RC is 6.55 m × 5.85 m ×

TABLE I. PARAMETERS OF V2X CHANNEL SOUNDER

| Key Parameters of the channel sounder | |
|---|---|
| Name of parameter (unit) | value |
| Centre Frequency | 3.5 GHz |
| Peak Output Power | 30 dBm |
| Ref accuracy | 1 ps |
| Symbols rate | 100 MS/s |
| PN sequency length | 4095 |
| Pluse shaping filter | 13 orders RRC filter roll off factor 0.25 |
| Max Capture length (cycles per snapshot) | 81920 (10 cycles per snapshot) |
| Antenna gain @3.5GHz | 6.8 dBi |
| Half Power Beamwidth @3.5GHz | E plan: 68 degree H plan: 80 degree |

3.5 m, it contains one vertically installed paddle stirrer. The transmit and receive antennas are the ETS-Lindgren's model 3117 double-ridged waveguide horn antennas, which are located at two corners of the RC three meters apart. The paddle stirrer is located between the two antennas and blocks the Line-of-Sight (LoS) path between them.

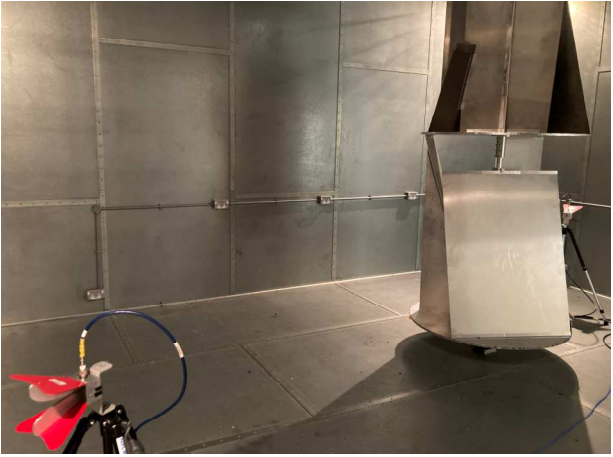

Fig. 4. RC measurement setups and the antenna pair

C. *Setup of channel emulator*

The CE used in the work was based on the vector signal transceiver PXIe-5644R which is set to 100 MHz channel sampling. Two typical channel models were selected for verification: 1) the Pedestrian-B model defined of in the 3GPP SCME channel model [6] which is a scenario with fewer multipaths; 2) the tapped delay line B (TDL-B) model with a 300 ns of delay spread scaling defined in the 3GPP 5G channel model [5] which is rich in multipaths. The two models are presented in the TABLE II and TABLE III. Limited by the SDR equipment minimum sample interval of 10 ns, several multipaths are indistinguishable and are merged in the channel emulator.

As shown in Fig. 5, the TDL-B model with rich multipaths and relatively small delay intervals is set in the CE. The Rayleigh distribution can be set for each path in terms of its LoS and non-line-of-sight (NLoS) settings. Doppler spread can also be imposed on each path according to Jakes spectrum model determined by the velocity of the motion experienced by the DUT.

TABLE II. PARAMETERS OF TAPS DEFINED IN PEDESTRIAN-B IN 3GPP 25.996

| Tap # | delay [ns] | Power [dB] | Fading distribution | Coerced delay* [ns] |
|---|---|---|---|---|
| 1 | 0.0000 | 0 | Rayleigh | 0 |
| 2 | 200 | -0.9 | Rayleigh | 200 |
| 3 | 800 | -4.9 | Rayleigh | 800 |
| 4 | 1200 | -8.0 | Rayleigh | 1200 |
| 5 | 2300 | -7.8 | Rayleigh | 2300 |
| 6 | 3700 | -23.9 | Rayleigh | 3700* |

*Note that due to the limitation of CE software, the last path at 3700 ns could not be set in the experiment.

TABLE III. PARAMETERS OF TAPS DEFINED IN TDL-B IN 3GPP 38.901

| Tap # | Normalized delay [ns] | Power [dB] | Fading distribution | Coerced delay* [ns] |
|---|---|---|---|---|
| 1 | 0 | 0 | Rayleigh | 0 |
| 2 | 0.03216 | -2.2 | Rayleigh | 30 |
| 3 | 0.06285 | -4 | Rayleigh | 60 |
| 4 | 0.06465 | -3.2 | Rayleigh | 60 |
| 5 | 0.0861 | -9.8 | Rayleigh | 90 |
| 6 | 0.08958 | -1.2 | Rayleigh | 90 |
| 7 | 0.11043 | -3.4 | Rayleigh | 110 |
| 8 | 0.11091 | -5.2 | Rayleigh | 110 |
| 9 | 0.11256 | -7.6 | Rayleigh | 110 |
| 10 | 0.15165 | -3 | Rayleigh | 150 |
| 11 | 0.15849 | -8.9 | Rayleigh | 160 |
| 12 | 0.171 | -9 | Rayleigh | 170 |
| 13 | 0.33063 | -4.8 | Rayleigh | 330 |
| 14 | 0.38268 | -5.7 | Rayleigh | 380 |
| 15 | 0.46422 | -7.5 | Rayleigh | 460 |
| 16 | 0.53526 | -1.9 | Rayleigh | 540 |
| 17 | 0.60507 | -7.6 | Rayleigh | 610 |
| 18 | 0.84882 | -12.2 | Rayleigh | 850 |
| 19 | 0.90657 | -9.8 | Rayleigh | 910 |
| 20 | 1.08561 | -11.4 | Rayleigh | 1090* |
| 21 | 1.23201 | -14.9 | Rayleigh | 1230* |
| 22 | 1.2837 | -9.2 | Rayleigh | 1280* |
| 23 | 1.43502 | -11.3 | Rayleigh | 1440* |

*Note that due to the limitation of CE software, the last 4 paths could not be set in the experiment.

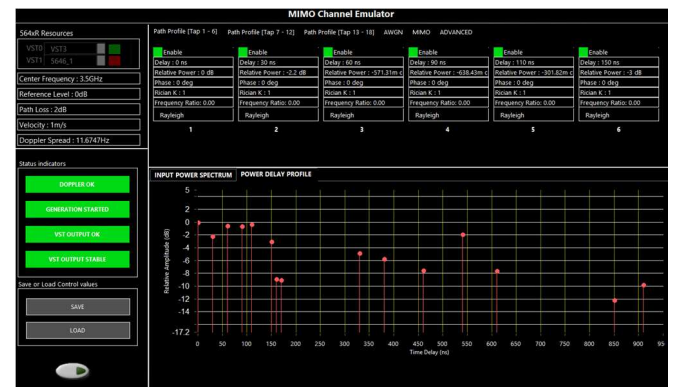

Fig. 5. the power delay profile setting of channel emulator

## D. Results

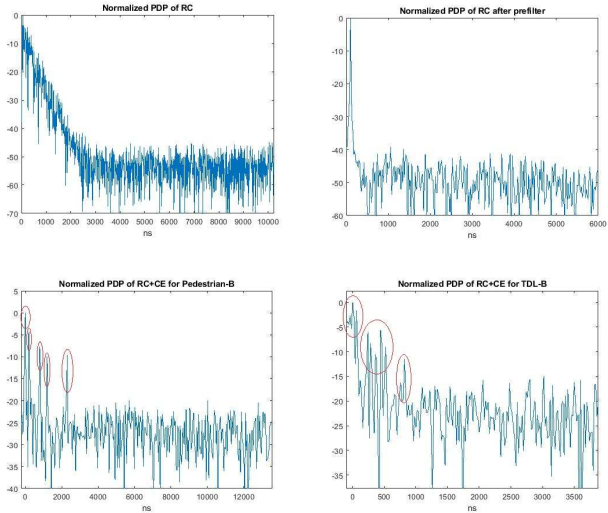

Fig. 6. (6a, Top left) the PDP of RC; (6b, Top right) the PDP of RC after convolving with the derived equalizer filter; (6c, Bottom left) the PDP of the synthesized channel for model Pedestrian-B; (6d, Bottom right) the PDP of the synthesized channel for model TDL-B with a 300 ns delay spread scaling.

In **Error! Reference source not found.**a, the PDP of RC at one stirrer position is presented which follows a typical continuous exponential delay profile with maximum delay value of approximately 2500 ns. After channel estimation and post-processing as described above, the equalizer filter was derived and output to the convolution module in the channel model synthesis step. To evaluate the effectiveness of the equalizer filter, a channel measurement which includes complete RF path in the channel model synthesis step and bypassing the CE is performed. From result of the Fig. 6b, it can be seen that one can achieve the promising cancellation effectiveness.

In **Error! Reference source not found.**c and **Error! Reference source not found.**d, the effectiveness of the two-step approach for two typical standard channel model is evaluated. In Fig. 6c, five discrete multipaths can be observed clearly with correct delay and power profile. Similar result could also be observed in simulation of rich multipaths scenario TDL-B of 3GPP 5G channel model from **Error! Reference source not found.**d, where clear discrete PDP can be seen and correspond correctly to the setting in the channel emulator.

## IV. Conclusion

This paper presents a novel two-step closed-loop method for accurate simulation of 3GPP 5G channel or SCME channels in an RC environment. From actual measurements in the RC, even if the channel sample rate of the CE is not very high, the efficiency of eliminating the inherent continuous channel response can be demonstrated. After evaluating the effectiveness of simulating two typical channel models in a RC environment, the method demonstrates the ability to extend the RC to performance tests defined in standards such as 3GPP, and the results can be further compared with other candidate 5G OTA methods such as MPAC and RTS.

## Acknowledgment

The work was supported in part by the 2021-2025 National Measurement System Programme of the UK government's Department for Business, Energy and Industrial Strategy (BEIS), under Science Theme Reference EMT23 of that Programme and in part by the EURAMET European Partnership on Metrology (EPM), under 21NRM03 Metrology for Emerging Wireless Standards (MEWS) project. The project (21NRM03 MEWS) has received funding from the EPM, co-financed from the European Union's Horizon Europe Research and Innovation Programme and by the Participating States.

## References

[1] 3GPP TR 37.977 V16.0.0 (2020-09), "Verification of radiated multi-antenna reception performance of User Equipment (UE)", Sept. 2020.

[2] M. Rumney, R. Pirkl, M. H. Landmann, and D. A. Sanches-Hernandez, "MIMO over-the-air research, development, and testing", Intenational Journal of Antennas and Propagation, 2012, Sept. 2012, pp. 1–8, doi:10.1155/2012/467695.

[3] CTIA, "Test Plan for 2x2 Downlink MIMO and Transmit Diversity Over-the-Air Performance", Version 1.2, Apr. 2018.

[4] D. Sanchez, M. A. Garciafernandez, N. Arsalane, M. Mouhamadou, D. Carsenat, and C. Decroze, "Validation of 3GPP SCME channel models emulated in mode-stirred reverberation chambers". 3GPP Tsg-ran Wg, May 2012.

[5] Chong Li, Tian-Hong Loh, ZhiHao Tian, Qian Xu, Yi Huang, "Evaluation of chamber effects on antenna efficiency measurements using non-reference antenna methods in two reverberation chambers", IET Microwaves, Antennas & Propagation, 09 August 2017

[6] 3GPP TR 38.901 V17.0.0 (2022-03), "Study on channel model for frequencies from 0.5 to 100 GHz", March. 2022

[7] Tien Manh Nguyen, Seung-Ho Kim, Jin-Young Jeong, Jong-Hwa Kwon, Jae-Young Chung, "Emulation of 3GPP SCME power-delay profiles for characterisation of multiple-input multiple-output antenna in reverberation chamber", IET Microwaves, Antennas & Propagation, 30 May 2018

[8] Xiaotao Guo, Tian Hong Loh, Yichi Zhang and Zhao He, "On the Realization Challenges for Accurate SCME Channel Implementation in RC", XXXIVth General Assembly and Scientific Symposium of the International Union of Radio Science (URSI GASS), 28 Aug. 2021

[9] 3GPP TS 25.996 V17.0.0 (2022-03), "Spatial channel model for Multiple Input Multiple Output (MIMO) simulations", March. 2022